\newcommand\revised[1]{\textcolor{black}{#1}}

\documentclass[sigconf]{acmart}

\AtBeginDocument{%
  \providecommand\BibTeX{{%
    \normalfont B\kern-0.5em{\scshape i\kern-0.25em b}\kern-0.8em\TeX}}}

\setcopyright{acmcopyright}
\copyrightyear{2018}
\acmYear{2018}
\acmDOI{10.1145/1122445.1122456}

\acmJournal{TOG}
\acmVolume{37}
\acmNumber{4}
\acmArticle{111}
\acmMonth{8}

\usepackage{amsmath}
\usepackage{listings}
\usepackage{xspace}
\usepackage{balance}
\usepackage{enumitem}
\usepackage{colortbl}
\usepackage{graphicx}
\usepackage{soul}
\usepackage{tcolorbox}
\usepackage{booktabs}
 
\usepackage{multicol}
\usepackage{multirow}
\usepackage[framemethod=tikz]{mdframed}
\usepackage{hyperref}
\usepackage{cleveref}

\definecolor{gray50}{gray}{.5}
\definecolor{gray40}{gray}{.6}
\definecolor{gray30}{gray}{.7}
\definecolor{gray20}{gray}{.8}
\definecolor{gray10}{gray}{.9}
\definecolor{gray05}{gray}{.95}

\usepackage{booktabs} 
\usepackage{framed}
\usepackage{graphicx}
\usepackage{subfigure}
\usepackage{pifont}
\usepackage{xspace}
\usepackage{multirow}
\usepackage{xcolor,colortbl}
\usepackage{hhline}
\usepackage{adjustbox}
\usepackage{booktabs} 
\usepackage{balance}
\usepackage{tcolorbox}
\usepackage{url}
\usepackage{booktabs}
\usepackage{paralist}
\usepackage{enumitem}
\usepackage{lipsum}
\usepackage{listings}
\usepackage[ruled,vlined,linesnumbered]{algorithm2e}
\definecolor{codegreen}{rgb}{0,0.6,0}
\definecolor{codegray}{rgb}{0.73,0.38,0.06}
\definecolor{codepurple}{rgb}{0.27,0.38,0.97}
\definecolor{codemagenta}{rgb}{0.74,0.09,0.42}
\definecolor{backcolour}{rgb}{0.96,0.96,0.96}

\usepackage[para]{footmisc}
\setcitestyle{numbers,sort&compress}

\newcommand\boost{\textsc{G-Mosa}\xspace}

\newcommand*{\ColorIfNotInString}[1]{\color{codegray}\textbf{#1}}%

\lstdefinestyle{mystyle}{
    backgroundcolor=\color{backcolour},   
    commentstyle=\color{codegreen},
    keywordstyle=\color{codepurple},
    numberstyle=\tiny\color{codegray},
    stringstyle=\color{codemagenta},
    language=Java,
    breakatwhitespace=false,         
    breaklines=true,                 
    keepspaces=true,                 
    numbers=left,
    xleftmargin=0.25cm,                    
    numbersep=5pt,                  
    showspaces=false,                
    showstringspaces=false,
    showtabs=false,                  
    tabsize=1,
    frame=tb,
    framerule=0pt,
    literate=%
    {0}{{{\ColorIfNotInString{0}}}}1
    {1}{{{\ColorIfNotInString{1}}}}1
    {2}{{{\ColorIfNotInString{2}}}}1
    {3}{{{\ColorIfNotInString{3}}}}1
    {4}{{{\ColorIfNotInString{4}}}}1
    {5}{{{\ColorIfNotInString{5}}}}1
    {6}{{{\ColorIfNotInString{6}}}}1
    {7}{{{\ColorIfNotInString{7}}}}1
    {8}{{{\ColorIfNotInString{8}}}}1
    {9}{{{\ColorIfNotInString{9}}}}1
}
\usepackage{tcolorbox}

\newcounter{Finding}
\stepcounter{Finding}

\usepackage{hyperref}
\usepackage{cleveref}
\RequirePackage{fontawesome}

\newcommand{\citeTodo}[1]{{\color{red}[??]}}

\newcommand{\ie}{\emph{i.e.},\xspace}

\newcommand{\eg}{\emph{e.g.},\xspace}

\newcommand{\etal}{\emph{et al.}\xspace}

{

\definecolor{gray50}{gray}{.5}
\definecolor{gray40}{gray}{.6}
\definecolor{gray30}{gray}{.7}
\definecolor{gray20}{gray}{.8}
\definecolor{gray10}{gray}{.9}
\definecolor{gray05}{gray}{.95}

\newlength\Linewidth
\def\findlength{\setlength\Linewidth\linewidth
  \addtolength\Linewidth{-4\fboxrule}
  \addtolength\Linewidth{-3\fboxsep}
}
\newenvironment{examplebox}{\par\begingroup
  \setlength{\fboxsep}{5pt}\findlength
  \setbox0=\vbox\bgroup\noindent
  \hsize=0.95\linewidth
  \begin{minipage}{0.95\linewidth}\normalsize}
  {\end{minipage}\egroup
  \textcolor{gray20}{\fboxsep1.5pt\fbox
    {\fboxsep5pt\colorbox{gray05}{\normalcolor\box0}}}
  \endgroup\par\noindent
  \normalcolor\ignorespacesafterend}

\newcommand\rqone{How does \boost compare to MOSA in terms of branch and mutation coverage?}

\newcommand\rqtwo{How does \boost compare to MOSA in terms of test case size?}

\newcommand\rqthree{How does \boost compare to MOSA in terms of maintainability of test cases?}

\newcommand\rqfour{How does \boost compare to MOSA in terms of understandability of test cases?}

\definecolor{gray50}{gray}{.5}
\definecolor{gray40}{gray}{.6}
\definecolor{gray30}{gray}{.7}
\definecolor{gray20}{gray}{.8}
\definecolor{gray10}{gray}{.9}
\definecolor{gray05}{gray}{.95}

\definecolor{deepskyblue}{rgb}{0.0, 0.75, 1.0}
\definecolor{arsenic}{rgb}{0.23, 0.27, 0.29}
\definecolor{emerald}{rgb}{0.31, 0.78, 0.47}
\definecolor{languidlavender}{rgb}{0.84, 0.79, 0.87}
\definecolor{magicmint}{rgb}{0.67, 0.94, 0.82}

\definecolor{arsenic}{rgb}{0.23, 0.27, 0.29}

\newtcolorbox{summarybox}[1]{
	colback=gray!9,
	colframe=black!,
	left=1.5mm,
	right=1.5mm,
	top=1.5mm,
	bottom=1.5mm,
	fonttitle=\bfseries,
	title=#1
}

\sloppy
\begin{document}

\title{Toward Granular Automatic Unit Test Case Generation}


\author{Fabiano Pecorelli}
\email{fabiano.pecorelli@tuni.fi}
\affiliation{%
  \institution{Tampere University}
  \city{Tampere}
  \country{Finland}
}

\author{Giovanni Grano}
\email{grano@ifi.uzh.ch}
\affiliation{%
  \institution{SEAL Lab - University of Zurich}
  \city{Zurich}
  \country{Switzerland}
}

\author{Fabio Palomba}
\email{fpalomba@unisa.it}
\affiliation{%
  \institution{SeSa Lab - University of Salerno}
  \city{Fisciano}
  \country{Italy}
}

\author{Harald C. Gall}
\email{harald.gall@uzh.ch}
\affiliation{%
  \institution{SEAL Lab - University of Zurich}
  \city{Zurich}
  \country{Switzerland}
}

\author{Andrea De Lucia}
\email{adelucia@unisa.it}
\affiliation{%
  \institution{SeSa Lab - University of Salerno}
  \city{Fisciano}
  \country{Italy}
}
\renewcommand{\shortauthors}{Pecorelli, et al.}

\begin{abstract}
  Unit testing verifies the presence of faults in individual software components. Previous research has been targeting the automatic generation of unit tests through the adoption of random or search-based algorithms. Despite their effectiveness, these approaches do not implement any strategy that allows them to create unit tests in a structured manner: indeed, they aim at creating tests by optimizing metrics like code coverage without ensuring that the resulting tests follow good design principles. In order to structure the automatic test case generation process, we propose a two-step systematic approach to the generation of unit tests: we first force search-based algorithms to create tests that cover individual methods of the production code, hence implementing the so-called \emph{intra-method tests}; then, we relax the constraints to enable the creation of \emph{intra-class tests} that target the interactions among production code methods. 
\end{abstract}

\setcopyright{acmcopyright}
\acmPrice{15.00}
\acmDOI{10.1145/xxx.xxx.xxx}
\acmYear{2022}
\copyrightyear{2022}
\acmISBN{xxx.xxx.xxx.xxx}
\acmConference[MSR'22]{Proceedings of the IEEE/ACM International Conference on Mining Software Repositories}{May 23-24, 2022}{Pittsburgh, PA, USA}
\acmBooktitle{Proceedings of the IEEE/ACM International Conference on Mining Software Repositories (MSR'22), May 23-24, 2022, Pittsburgh, PA, USA}

\ccsdesc[500]{Software and its engineering~Search-based software engineering}
\ccsdesc[500]{Software and its engineering~Empirical software validation}

\keywords{Search-based Software Testing, Test Code Quality, Automatic Test Case Generation}

\maketitle

    \section{Introduction}
\label{sec:introduction}

Software testing is the process adopted by developers to verify the presence of faults in production code \cite{myers2011art}. The first step of this process consists of assessing the quality of individual production code units \cite{ammann2016introduction}, \eg classes of an Object-Oriented project. Previous studies \cite{erdogmus2005effectiveness,williams2009effectiveness} have shown that unit testing alone may identify up to 20\% of a project's defects and reduce up to 30\% the costs connected with development time. Despite the undoubted advantages given by unit testing, things are worse in reality: most developers do not actually practice testing and tend to over-estimate the time spent in writing, maintaining, and evolving unit tests, especially when it comes to regression testing \cite{beller2017developer}.

To support developers during unit testing activities, the research community has been developing automated mechanisms---relying on various methodologies like random or search-based software testing \cite{anand2013orchestrated}---that aim at generating regression test suites targeting individual units of production code. For instance, Fraser and Arcuri \cite{fraser2013whole} proposed a search-based technique, implemented in the \textsc{Evosuite} toolkit,\footnote{http://www.evosuite.org} able to optimize whole test suites based on the coverage achievable on production code by tests belonging to the suite. Later on, Panichella \etal \cite{panichella15reformulating} built on top of \textsc{Evosuite} to represent the search process in a multi-objective, dynamic fashion that allowed them to outperform the state-of-the-art approaches. Further techniques in literature proposed to (1) optimize code coverage along with other secondary objectives (\ie performance \cite{ferrer2012evolutionary,grano2019testing,pinto2010multi}, code metrics \cite{oster2006automatic,palomba2016automatic}, and others \cite{lakhotia2007multi}) or (2) empower the underlying search-based algorithms by working on their configuration \cite{arcuri2019restful,knowles2000approximating,zamani2020cost}. Yet, these approaches often fail to generate tests that are well-designed, easily understandable, and maintainable \cite{fraser2013whole}.

\revised{In addition, existing approaches do not explicitly follow well-established methodologies that suggest taking \emph{test case granularity} into account \cite{pezze2008software}. In particular, when developing unit test suites, two levels of granularity should be preserved \cite{harrold1992incremental,orso1998open,pezze2008software}: first, the creation of tests covering single methods of the production code should be pursued, \ie \emph{intra-method} \cite{pezze2008software} or \emph{basic-unit} testing \cite{orso1998open}; afterwards, tests exercising the interaction between methods of the class should be developed in order to verify additional execution paths of the production code that would not be covered otherwise, \ie \emph{intra-class} \cite{pezze2008software} or \emph{unit} testing \cite{orso1998open}. Besides producing test cases of higher quality, a structured strategy might potentially lead to the generation of tests whose oracle would be easier to be find for developers, as they would be required to check smaller portions of code to identify the expected behavior \cite{barr2015oracle}.}

\revised{In this paper, we target the problem of granularity in automatic test case generation, advancing the state of the art by pursuing the first steps toward the integration of a systematic strategy within the inner-working of automatic test case generation approaches that might possibly support the production of more effective and understandable test suites.} We build on top of \textsc{Mosa} \cite{panichella15reformulating} to devise an improved technique, coined \textsc{Granular-Mosa} (\boost hereafter), that implements the concepts of intra-method and intra-class testing. Our technique splits the overall search budget in two. In the first half, \boost forces the search-based algorithm to generate intra-method tests by limiting the number of production calls to one. In the second half, the standard \textsc{Mosa} implementation is executed so that the generation can cover an arbitrary number of production methods, hence producing intra-class test cases that exercise the interaction among methods.



\revised{We conjecture that, given the same amount of time, it could be possible to improve the methodology implemented to automatically generate unit tests in order to produce test suites with different granularity levels, following a structured procedure.}


	
	

\smallskip 
\noindent \textbf{Structure of the paper.} Section \ref{sec:background} provides background required to properly understand our research. In Section \ref{sec:approach} we present the algorithmic details of \textsc{G-Mosa}, while Section \ref{sec:RQs} overviews the research questions that we will address. In Section \ref{sec:emp-study} we report on the experimental plan of the evaluation of our technique. Section \ref{sec:limitations} discusses the possible limitations of both the approach and the experimental plan. Finally, Section \ref{sec:conclusions} outlines our next steps.

    \section{Background}
\label{sec:background}
This section reports the basic concepts on automated tools to generate unit test suites as well as a discussion on related work.

\subsection{Automatic Unit Test Case Generation}
The problem of automatically generating test data has been largely investigated in the last decade~\cite{mcminn2004search-based}.
Search-based heuristics---genetic algorithms~\cite{goldberg2006genetic} in particular---have been successfully applied to solve such a problem~\cite{mcminn2004search-based} with the goal to generate tests with high code coverage.
Single-target approaches have been the first techniques proposed in the context of white-box testing~\cite{scalabrino2016search-based}.
These approaches divide the search budget among all the targets (typically branches) and attempt to cover each of them at a time. 
To overcome the limitation of single-target approaches, Fraser and Arcuri~\cite{fraser2013whole} proposed a multi-target approach, called \emph{whole suite test generation (WS)}, that tackles all the coverage targets at the same time.
Building on such idea, Panichella \etal~\cite{panichella15reformulating} proposed a many-objective algorithm called MOSA.
While WS is guided by an aggregate suite-level fitness function, MOSA evaluates the overall fitness of a test suite based on a vector of \emph{n} objectives, one for each branch to cover.
The basic working of MOSA can be summarized as follows.
At first, an initial population of randomly generated tests is initialized.
Such a population is then evolved through consecutive generations:
new offsprings are generated by selecting two parents in the current population and then both crossover and mutation operators are applied~\cite{panichella15reformulating}.
MOSA introduced a novel \emph{preference-sorting} algorithm to focus the search toward uncovered branches. 
This heuristic solves the problem of selecting non-dominated solutions that typically occurs in many-objective algorithms~\cite{von2014survey}. 

\paragraph{Random Test Case Generation}
\begin{algorithm}[!th]
  \DontPrintSemicolon
  \scriptsize
  \SetAlgoLined
  \SetAlgoVlined
  \SetKwInOut{Input}{Input}
  \Input{$M=\{m_1,m_2,...,m_i\}$: methods of the CUT we want to cover\\
  	\indent Maximum attempts $A$ \\
  	\indent Maximum size $L$\\
  }
  \KwResult{$T\{s_1,s_2,...,s_n\}$: test case with with $n$ statements}
   \Begin{
   		$T \leftarrow \emptyset$\;
   		$r \leftarrow$ RANDOM-NUMBER(1, $L$)\;
   		\While{not(max attempts reached) AND ($|T|\leq$ L)}{
   			$p \leftarrow$ RANDOM-NUMBER(0, 1)\;
   			\eIf{$p \leq$ INSERTION-UUT}{
   				INSERT-CALL-ON-CUT($T$)\;
   			}
   			{
	            $v \leftarrow$ SELECT-VALUE($T$)\;
    	        INSERT-CALL-ON-VALUE($T$, $v$)\;
   			}
   		}
  		\Return{$T$}\;
  }
  \caption{Random Test Cases Generation}
  \label{algo:generation}
\end{algorithm}

To provide the reader with the necessary context, we introduce the basics of the mechanism used by EvoSuite~\cite{evosuite} to randomly initialize the first generation of tests.
More details can be found in the paper by Fraser and Arcuri~\cite{fraser2013whole}.
A tests case is represented in EvoSuite by a sequence of statements $T = \{s_1, s_2, ..., s_l\}$ where $|T| = l$.
Each $s_i$ has a particular value $v(s_i)$ of type $\tau$.
The pseudo-code for the random test cases generation is showed in Algorithm~\ref{algo:generation}.
At first, EvoSuite chooses a random $r \in (1, L)$ where $L$ is the test maximum length (\ie number of statements) (line 3 of Algorithm~\ref{algo:generation}).
Thus, EvoSuite initializes an empty test and tries to add new statements to it.
Such a logic is implemented in the \texttt{RandomLengthTestFactory} class.
EvoSuite defines five different kinds of statements \cite{fraser2013whole}: 
(i) primitive statements ($S_p$), \eg creating an Integer or a String variable,
(ii) constructor statements ($S_c$), that instantiate an object of a given type,
(iii) field statements ($S_f$) that access public member variables,
(iv) method statements ($S_m$), \ie method invocations on objects (or static method calls), and
(v) assignment statements ($S_a$) that assign a value to a defined variable.
The value $v$ and the type $\tau$ of each statement depend on the generated statement itself, \eg the value and type of method statement will depend on the return value of the invoked method.
In the preprocessing phase, a test cluster~\cite{wappler2005using} analyzes the entire SUT (\emph{system under test}) and identifies all the available classes $\Omega$.
$\forall c \in \Omega$, the test cluster defines a set of $\{\mathcal{C}, \mathcal{M}, \mathcal{F}\}$, where respectively, $\mathcal{C}$ is the set of constructors, $\mathcal{M}$ if the set of instance method and $\mathcal{F}$ is the set of instance fields available for a class $c$.
 
EvoSuite tries to repetitively generate new statements (the loop from line 4 to line 10 in Algorithm~\ref{algo:generation}) and add them to a test.
The process continues until the test hits the maximum random length or the maximum number of attempts (a parameter in EvoSuite set to 1,000 by default) is reached (line 4 in Algorithm~\ref{algo:generation}).
EvoSuite can insert two main kinds of statements.
With a probability lower than \textsc{INSERTION-UUT} (a property defined to 0.5 by default), EvoSuite generates a random call of either a constructor of the class under test (CUT) or a member class, \ie instance field of method (lines 6-7 in Algorithm~\ref{algo:generation}).
Alternatively, the tool can generate a method call to a value $v(s_j)$ where $j \in (0, i]$ and $i$ is the position on which the statements will be added (lines 9-10 in Algorithm~\ref{algo:generation}). 
In other words, EvoSuite invokes a method on a value of a statement already inserted into the test. 
Such a value is randomly selected among all the values from the statements from the position 0 to the actual position (line 9 in Algorithm~\ref{algo:generation})
EvoSuite also takes care of the parameters or the callee objects needed to generate a given statement.
For example, a call to an instance method of the CUT requires 
(i) the generation of a statement instantiating the CUT itself, and 
(ii) the generation of a statement defining values needed as argument for the method call.
The values for such parameters can either
(i) be selected among the sets of values already in the test,
(ii) set to \texttt{null}, or 
(iii) generated randomly.
 
\begin{lstlisting}[caption={Example of a test generated by Evosuite}, captionpos=b, label={code:example},float={h}]
StringReader stringReader0 = new StringReader("#<z-K~+*O4@s^W");
char[] charArray0 = new char[1];
stringReader0.read(charArray0);
JavaCharStream javaCharStream0 = new JavaCharStream(stringReader0, (-1273), 1, 77);
JavaParserTokenManager javaParserTokenManager0 = new JavaParserTokenManager(javaCharStream0);
Token token0 = javaParserTokenManager0.getNextToken();
\end{lstlisting}
 
To better understand the generation process, let consider the test case in Listing~\ref{code:example}, which has been generated for the class \texttt{JavaParserTokenManager}.
To create this test, Evosuite works as follows.
Starting from an empty test, it decides with a certain random probability to insert a statement invoking an instance method of the CUT: in our example, the \texttt{getNextToken()} method (line 6 of Listing~\ref{code:example}).
However, Evosuite needs first to generate two other statements, \ie line 5 and 6 of Listing~\ref{code:example}, respectively:
a statements returning a value of type \texttt{JavaParserTokenManager} (\ie the callee of the method) and a statement returning a value of type \texttt{JavaCharStream} (\ie the parameter of the method).
In turn the constructor of \texttt{JavaCharStream} will need a value of type \texttt{StringReader} (line 1 of Listing~\ref{code:example}).
Line 3 of Listing~\ref{code:example} is instead the result of the other kind of possible insertion, \ie a method call to a value already present in the test: the \texttt{stringReader0} object in this case.
Similarly, the tool will generate the primitive statement at line 2 of Listing~\ref{code:example} to provide the parameter needed by such a call.

    \section{\boost: A Two-Step Automatic Test Case Generation Approach}
\label{sec:approach}
\begin{algorithm}[!tb]
  \DontPrintSemicolon
  \scriptsize
  \SetAlgoLined
  \SetAlgoVlined
  \SetKwInOut{Input}{Input}
  \Input{$B=\{\tau_1,...,\tau_m\}$: set of coverage targets of a program\\
  \indent Population size $M$\\
  }
  \KwResult{A test suite $T$}
   \Begin{
		$T \leftarrow \emptyset$\;
		$\alpha \leftarrow$ intra-method-testing($M$)\;
		$\gamma \leftarrow$ MOSA($M$)\;
		$T_\alpha, B_\alpha \leftarrow$ GENERATE-TESTS($\alpha, B$) \tcc*[r]{half search budget}
		\If{$B_\alpha == \emptyset$} {
			\Return{$T_\alpha$}\;
		}
		$T \leftarrow T \bigcup T_\alpha$\;
		$T_\gamma, B_\gamma \leftarrow$ GENERATE-TESTS($\gamma, B_\gamma$) \tcc*[r]{half search budget}
		$T \leftarrow T \bigcup T_\gamma$\;
  		\Return{$T$} \;
  }
  \caption{\boost Algorithm}
  \label{algo:main:approach}
\end{algorithm}
\boost is defined as a two-step methodology that combines intra-method and intra-class unit testing~\cite{orso1998open,pezze2008software}.
The pseudo-code of \boost is outlined in Algorithm~\ref{algo:main:approach}.
The first step of the methodology generates tests that exercise the behavior of production methods in isolation: we indeed only allowed by design to generate intra-method tests (details in~\Cref{subsec:bumosa}).
The second step is based on the standard MOSA implementation~\cite{panichella15reformulating} that performs intra-class unit testing by exercising a class trough a sequence of method call invocations.
In the following, we detail each of these two steps.

\subsection{Step I - Intra-Method Tests Generation}
\label{subsec:bumosa}
\begin{algorithm}[!tb]
  \DontPrintSemicolon
  \scriptsize
  \SetAlgoLined
  \SetAlgoVlined
  \SetKwInOut{Input}{Input}
  \Input{$T\{s_1,s_2,...,s_n\}$: test case with with $n$ statements\\
  	\indent $S=\{s_1,s_2,...,s_j\}$: setters of the CUT\\
  }
  \KwResult{$T$: test case with with $n+1$ statements}
   \Begin{
   		$o \leftarrow $ GET-RANDOM-TEST-CALL\;
   		\eIf{($o$ is a method)}{
   			\If{RANDOM-NUMBER(0, 1) $\leq$ INSERTION-SET}{
   				$T \leftarrow T \bigcup$ ADD-METHOD($s_j \in S$)\;
   				\Return{$T$}\; 
   			}
   			$T \leftarrow T \bigcup$ ADD-METHOD($o$)\;
   			$T_c \leftarrow$ true
   		}
   		{
   			$T \leftarrow T \bigcup$ (ADD-COSTRUCTOR($o$) OR $T \bigcup$ ADD-FIELD($o$))\;
   		}
  		\Return{$T$}\;
  }
  \caption{Insert Random Call}
  \label{algo:random:call}
\end{algorithm}
The intra-method testing process is the first step to be initialized (line 3 of Algorithm~\ref{algo:main:approach}).
Like any other test-case generation technique, a set of coverage targets $B$ is given as input.
The intra-method process starts (line 5 of Algorithm~\ref{algo:main:approach}) with $B$ as target of the search and sets its search budget to the half of the overall budget available:
in other words, if \boost is given 180 seconds as budget, the intra-method testing process will run for 90 seconds.
At the end of its search, the first step returns
(i) $T_\alpha$, the set of generated tests cases, and 
(ii) $B_\alpha$, the set of uncovered targets.
$T_\alpha$ and $B_\alpha$ will be used then as input for the second phase (see ~\Cref{subsec:mosa}).

\paragraph*{Intra-Method Code-Generation Engine}
As already mentioned, \boost is a variant of \textsc{Mosa} that applies first an intra-method testing methodology~\cite{orso1998open}:
each generated test exercises a single production method of the CUT.
To enable intra-method testing, we modified the code-generation engine used by \textsc{EvoSuite} to randomly generate new tests.
In~\Cref{sec:background} we described such a mechanism: 
in a nutshell, \textsc{EvoSuite} inserts randomly generated statements (\eg calls to a class constructor or invocation of instance methods) in a test until a maximum number of statements is reached.
This approach does not guarantee---nor has been designed to do it---any control on the number of instance method invocations of a test.
As a consequence, tests might end up containing a sequence of method calls for the CUT and thus, perform intra-class unit testing. 

To enable intra-method testing, we modified the high-level algorithm described in Algorithm~\ref{algo:generation}.
With the current formulation, the insertion loop (from line 4 to line 10 in Algorithm~\ref{algo:generation}) has two stopping conditions:
either a maximum number of attempts or the maximum length $L$ of the test is reached. 
We defined a third stopping criteria: 
as soon as a statement $s_i$ representing a method invocation on a CUT object is inserted, we considered the test as complete.
To store this information, in our implementation each test $T$ has a property $T_c$, initially set to \texttt{false}, that indicates whether such a statement $s_i$ has been inserted in $T$.
Therefore, we added $not(T_c)$ as additional stopping criteria for the insertion loop at line 4 of Algorithm~\ref{algo:generation}.
It is worth remarking that insertions of CUT instance methods are managed by the INSERT-CALL-ON-CUT procedure (line 7 of Algorithm~\ref{algo:generation}).
Thus, we re-implemented such a procedure to handle the newly defined stopping criteria.

Algorithm~\ref{algo:random:call} shows our ad-hoc implementation of the INSERT-CALL-ON-CUT procedure.
The algorithm takes as input a test $T$ with $1 \leq n < L$ statements and a set $S \subseteq \langle \mathcal{M}_{CUT} \cup \mathcal{F}_{CUT} \rangle$ of setters for the CUT.
For a class $c$, $S$ is composed of all its instance fields $\mathcal{F}$ and of a subset of its instance methods $\mathcal{M}$.
We defined the following heuristic to detect the instance method $\in S$ for the CUT.
We considered as setter every $m \in \mathcal{M}$ whose method name has the $\langle \textrm{prefix} \rangle \langle \textrm{keyword} \rangle \langle \textrm{suffix} \rangle$ structure, with $\textrm{keyword} \in \{\textrm{set}, \textrm{get}, \textrm{put}\}$, if and only if 
$\exists \; m' \in M \; | \; \langle \textrm{keyword}\rangle' == \textrm{get}$ 
and $\langle \textrm{prefix}\rangle' == \langle \textrm{prefix} \rangle \; \& \; \langle \textrm{suffix}\rangle' == \langle\textrm{suffix}\rangle$.
It is worth noting that the $\langle prefix \rangle$ part of the method name is optional.
For instance, let consider the class \texttt{SimpleNode} of the \textsc{jmca} project: this has two instance methods named \texttt{jjtSetParent} and \texttt{jjtGetParent}.
According to our heuristic, the method \texttt{jjtSetParent} is considered as a setter method of the class \texttt{SimpleNode}.

The first step for generating a random call on the CUT is to extract a random call $o$ in the set $\{\mathcal{C}, \mathcal{M}, \mathcal{F}\}$.
This is done by the GET-RANDOM-TEST-CALL procedure (line 2 of Algorithm~\ref{algo:random:call}).
If $o \in \{\mathcal{C} \cup \mathcal{F}\}$, a new statement $s_i$ including a call to $o$ is inserted into the test (as described in~\Cref{sec:background}).
In case $o \in M$---with a certain probability (set as property to 0.3 by default)---a new statement with a randomly selected setter is generated and inserted into $T$; therefore, the test is returned (lines from 4 to 6 in Algorithm~\ref{algo:random:call}).
In the opposite case, $o$ is added to the test $T$ and its property $T_c$ is set to \texttt{true} (lines 7 and 8 of Algorithm~\ref{algo:random:call}).
As a consequence, the code-generation engine stops attempting new insertions: $T_c$ is now \texttt{true} and the condition $not(T_c)$ is not met anymore.
Our implementation of GET-RANDOM-TEST-CALL enables intra-method testing since it allows by design the invocation of a single instance method of the CUT.
Note that our formulation does not consider setters as units under test since they are needed only to set the state of the CUT object required to properly exercise the method under test.

\subsection{Step II - Intra-Class Tests Generation}
\label{subsec:mosa}
As previously explained, the GENERATE-TESTS procedure returns a set of generated tests ($T_\alpha$) and a set of uncovered targets $B_\alpha \subseteq B$.
If $B_\alpha == \emptyset$, the intra-method testing process achieved full coverage on the CUT and $T_\alpha$ is returned (lines 6-7 of Algorithm~\ref{algo:main:approach}).
In the opposite case, $T_\alpha$ is added to $T$ (line 8 of Algorithm~\ref{algo:main:approach}).
In the second step, \textsc{Mosa} is selected as algorithm for the search.
This time, $B_\alpha$ is given as set of target to \textsc{Mosa} (line 9 of Algorithm~\ref{algo:main:approach}).
In other words, \textsc{Mosa} will attempt to cover only what has not been covered in the first step.
At the end of the GENERATE-TESTS procedure, the resulting $T_\gamma$ is added to $T$ and the final test suite $T$ is returned (lines 9-10 of Algorithm~\ref{algo:main:approach}).
$T$ is formed by two different kinds of tests: 
$T_\alpha$ generated by the intra-method process, that tests single production methods in isolation and 
$T_\gamma$ generated by \textsc{Mosa}, that exercise a class by constructing sequences of method calls.
 
    \section{Research Questions and Objectives}
\label{sec:RQs}
The \emph{goal} of the empirical study, according to the Goal-Question-Metric (GQM) template \cite{caldiera1994goal}, is: \emph{evaluate} the effectiveness, size, and maintainability of the test suites generated by \boost with the \emph{purpose} of understanding the extent to which our approach can generate higher-quality unit test cases when compared to a state of the art automatic test case generation technique like \textsc{Mosa}. 

To address our goal, we set up three \emph{research questions} (\textbf{RQ}s). 
In the first place, we target one of the risks associated with the mechanisms implemented in \boost, namely the decrease of both code and mutation coverage: by design, we force \boost to generate intra-method tests, naturally limiting its scope and potentially lowering the number of tangentially covered branches. 

Our first \textbf{RQ} can therefore be seen as preliminary and aimed at assessing this aspect by comparing the effectiveness of test suites generated by \boost and \textsc{Mosa}~\cite{panichella15reformulating}. We consider \textsc{Mosa} as baseline because (1) previous techniques aimed at improving the quality of generated tests were compared to \textsc{Mosa} as well (\eg ~\cite{palomba2016automatic}) and (2) we built \boost on top of \textsc{Mosa}, making the comparison required. We define the following research question:

\begin{center}
	\begin{examplebox}
		\textbf{RQ$_1$ - Effectiveness.} \emph{\rqone}	
	\end{examplebox}
\end{center}

Once assessed the implications of \boost for the effectiveness of test cases, we aim to investigate the potential benefits given by our technique. \revised{We take into account the size of the generated test cases: according to previous research in the field \cite{panichella15reformulating,fraser2013whole,grano2020pizza}, this is an indicator that has been often used to estimate the effort that developers would spend to comprehend and interact with the tests, indeed, a number of previously proposed search-based automatic test case generation approaches used it as a metric to optimize \cite{Panichella:2015we,oster2006automatic,pinto2010multi}.} Also in this case, we compare the size of test cases generated by \boost and \textsc{Mosa}, addressing the following \textbf{RQ}:


\begin{center}
	\begin{examplebox}
		\textbf{RQ$_2$ - Size.} \emph{\rqtwo}
	\end{examplebox}
\end{center}

\revised{While the size assessment could already provide insights into the comprehensibility of the generated test cases, in the context of our research we aim to provide additional analyses to assess their potential usefulness from a maintainability perspective. In particular, once generated, test cases not only need to be manually validated by testers to verify assertions \cite{afshan2013evolving,barr2015oracle}, but also maintained to keep them updated as a consequence of the changes to the production code \cite{palomba2016automatic}. Hence, it is reasonable to assess the capabilities of our approach in this respect. We plan to compare \boost and \textsc{Mosa} in terms of the metrics that have been previously designed to describe the quality and maintainability of test cases and that we have surveyed in our previous work \cite{pecorelli2021relation}. These pertain to (1) code complexity, as measured by the weighted method count of a test suite \cite{subramanyam2003empirical}; (2) efferent coupling \cite{fregnan2019survey}; and (3) test smells, i.e., suboptimal design or implementation choices applied when developing test cases \cite{garousi2018smells}. 
This lead to our third research question:}

\begin{center}
	\begin{examplebox}
		\textbf{RQ$_3$ - Maintainability.} \emph{\rqthree}
	\end{examplebox}
\end{center}

\revised{On the one hand, the quantitative measurements computed so far can provide a multifaceted view of how the proposed approach compares to state of the art in terms of performance. On the other hand, these analyses cannot quantify the actual gain given by \textsc{G-Mosa} in practice. For this reason, the last step of our methodology includes a user study where we plan to inquiry developers on the understandability of the test cases output by \textsc{G-Mosa} when compared to those of \textsc{Mosa}. This leads to the formulation of our last research question:}

\begin{center}
	\begin{examplebox}
		\revised{\textbf{RQ$_4$ - Understandability.} \emph{\rqfour}}
	\end{examplebox}
\end{center}

    \section{Experimental plan}
\label{sec:emp-study}
To answer our research questions, we aim to perform an empirical study on Java classes comparing \boost to MOSA~\cite{panichella15reformulating}. This section reports details about the experimental procedure planned to address our RQs. 

\subsection{Experimental Environment}
We will run \boost and \textsc{Mosa} against a dataset of Java classes, collecting the generated tests and the corresponding code coverage indicators. In particular, we plan to consider around 100 classes pertaining to the SF110 corpus~\cite{fraser2014large}. This benchmark\footnote{\url{http://www.evosuite.org/experimental-data/sf110/}} contains a set of Java classes extracted from 110 projects of the \textsc{SourceForge} repository. We select it since this is typically used in automatic test case generation research \cite{fraser2014large,panichella15reformulating,grano2019scented,fraser2013whole} and, therefore, can allow us to experiment our technique on a ``standard'' benchmark that would enable other researchers to build upon our findings and compare other techniques. 

To account for the intrinsic non-deterministic nature of genetic algorithms, we will run each approach on each class in the dataset for 30 times, as recommended by Campos \etal~\cite{campos2017empirical}.
We use the time criterion as search budget, allowing 180 seconds for the search~\cite{campos2017empirical}. In \boost, this time is equally distributed amongst the two steps of the approach, \ie we reserve 90 seconds for intra-method and 90 for intra-class testing. \textsc{Mosa} could instead rely on the entire search budget to generate tests, as it does not have multiple steps.

To run the experimented approaches, we rely on the default parameter configuration given by \textsc{Evosuite}. As shown by Arcuri and Fraser \cite{arcuri2013parameters}, the parameter tuning process is long and expensive, other than not necessarily paying off in the end.

\subsection{Collecting Performance Metrics} 
In the context of \textbf{RQ$_1$}, we will rely on the code and mutation coverage analysis engine of \textsc{Evosuite}~\cite{fraser2015achieving}. In particular, we will let the tool collect the branch coverage of each test in each of the 30 runs. Additionally, the tool will also collect information on the mutation score: despite the existence of other tools able to perform mutation analysis (\eg \textsc{PiTest}\footnote{The \textsc{Pitest} analyzer: \url{https://pitest.org}.}), we rely on the one made by \textsc{Evosuite} since it can effectively represent real defects~\cite{fraser2015achieving} and has been used in a series of recent studies on automatic test case generation~\cite{grano2019testing, panichella2018automated, panichella2018incremental}. We aim to perform the mutation analysis at the end of the search, once generated the unit tests for all the approaches. 
To obtain meaningful results we give an extra-budget of 5 minutes to the mutation analysis---this step is required to generate more mutants and to verify the ability of tests to capture them~\cite{fraser2015achieving}. 

As for \textbf{RQ$_2$}, we start from the set of test suites output by the search process for the two experimented approaches and first compute their overall size, \ie the lines of code of the generated test classes. As shown by previous work in the field \cite{fraser2013whole,panichella2018automated}, this metric represents an indicator of the usability of the test suites given by the tools. While recognizing the value of this perspective, we also know that such a validation could be excessively unfair in our case. By design, \boost aims at creating a larger amount of test cases with respect to \textsc{Mosa}, with a first set of many small tests implementing the concept of intra-method testing and a second set composed of larger tests that implement the concept of intra-class testing. On the contrary, \textsc{Mosa} does not explicitly target the creation of maintainable test cases, hence possibly generating a fewer amount of tests that account for a lower overall test suite size while reaching high branch coverage. As a consequence, the assessment of the overall test suite size could be too simplistic, other than providing coarse-grained considerations on the usefulness of test suites, \ie in practice, developers rarely look at the \emph{entire} test suite while fixing defects \cite{ceccato2015automatically}. Hence, we aim to complement the overall test suite size assessment with an analysis of the properties of the individual test cases: we plan to compute the \emph{mean size per test case}, namely the average amount of lines of code of the automatically generated test cases within a test suite. Such a measurement can allow us to verify whether our approach could provide developers with smaller units that might better align to the actual effort required by a developer to deal with the tests generated by \boost when compared to our baseline \textsc{Mosa} \cite{ceccato2015automatically}. 

\revised{To answer our third research question (\textbf{RQ$_3$}), we aim to compute three metrics which have been previously associated with maintainability and that might affect the way developers interact with test cases \cite{spadini2018relation,pecorelli2021relation,grano2020pizza,gren2017relation}. Weighted Method Count of a Test Suite (TWMC) \cite{subramanyam2003empirical} represents a complexity metric whose computation implies the sum of the complexity values of the individual test cases of a test suite. The metric provides an estimation of how complex a test would be to understand for a developer \cite{elish2006design,gren2017relation}. In the second place, we compute the efferent coupling metric (EC) \cite{fregnan2019survey}, which provides an estimation of how coupled the test cases of a suite are with the other test cases of the same suite. Keeping coupling under control is a key concern when writing test cases, as an excessive dependence among tests might potentially lead to some sort of flakiness \cite{habchi2021qualitative}. Finally, we will detect the number of test smells per test suite: these smells have been often associated to the decrease of maintainability and effectiveness of test suites \cite{spadini2018relation,grano2019scented} and likely represent the most suitable maintainability aspect to verify within the test code. In this respect, it is worth remarking that automatically generated test code is by design affected by certain test smells: for instance, the generated tests come without assertion messages and, therefore, are naturally affected by the smell known as \emph{Assertion Roulette} \cite{garousi2018smells}, which arises when a test has no documented assertions. At the same time, automatic tests might not suffer from other types of smells. For example, external resources are mocked by the \textsc{Evosuite} framework, making the emergence of a test smell like \emph{Mystery Guest} \cite{garousi2018smells}---which has to do with the use of external resources---not possible. As such, comparing the experimented approaches based on the presence of these smells would not make sense. Hence, we will only consider the test smells whose presence can be actually measured.}
\revised{In more practical terms, we will employ the tool by Spinellis \cite{spinellis2005tool} to compute TWMC and EC metrics. As for test smells, we will rely on \textsc{TsDetect} \cite{peruma2020tsdetect}, which is a tool able to identify more than 25 different types of test smells---in this case, however, we will limit the detection to the test smells that might actually arise in automatically generated tests.}

\revised{\subsection{Collecting Understandability Metrics}}
\revised{The last step of our experimentation concerns with the assessment of the actual gain provided by \textsc{G-Mosa} in practice. We therefore design an online experiment where we will (1) involve developers in tasks connected to the understandability of the test cases generated by our approach and (2) compare our approach with the baseline \textsc{Mosa} implementation. Our experiment will be online as we are not currently allowed to perform in-situ controlled experiments because of the COVID19 pandemic. We will therefore use an online platform we have recently developed and that allow external participants to (1) navigate and interact with source code elements and (2) answer closed and open questions. 
}

\smallskip \revised{\textbf{Participant's recruitment.} We will recruit developers using various channels. In the first place, we will invite the original open-source developers of the classes considered in the study. This will be done via e-mail. Of course, we will only approach the developers who have publicly released their e-mail address on \textsc{GitHub}. In a complementary manner, we will recruit participants through \textsc{Prolific}\footnote{\textsc{Prolific} website: \url{https://www.prolific.co/}.} by carefully considering the guidelines recently proposed by Reid et al. \cite{reid2022_prolific_recommendations}. This is a research-oriented web-based platform that enables researchers to find participants for user studies. One of the features of \textsc{Prolific} is the specification of constraints over participants, which in our case enabled to limit the participation to software developers. It is important to point out that \textsc{Prolific} implements an \emph{opt-in} strategy \cite{hunt2013participant}, meaning that participants get voluntarily involved. This might potentially lead to self-selection or voluntary response bias \cite{heckman1990selection}. To mitigate this risk, we will introduce an incentive of 7 pounds per valid respondent. Once we will receive the answers, we will filter out the answers coming from developers without the minimum expertise required (e.g., developers without any expertise on testing) and participants who did not take the task seriously. Should the amount of participants not reached an acceptable level, quantified in 120 valid participants, we will conduct the recruitment again until reaching the acceptable bar.}




\smallskip
\revised{\textbf{Experimental setting.}}
\revised{The participants will be first asked to answer demographic questions that will serve to address their background and level of expertise in software development and testing.}

\revised{Then, participants will be asked to perform the same task twice. In particular, they will be provided with the source code of two \textsc{Java} test classes aiming at exercising the same production class. One of them will be generated by \boost and the other one by \textsc{Mosa}. In each task, after reading each of the two test classes participants will be asked to (1) rate the overall understandability of the class with a 5-points Likert scale (from 1, which indicates poorly understandable code, to 5, which indicates fully understandable code); (2) explain the reasons for the rating provided; (3) Write the assertion messages for the methods of the test class under consideration. With the first two questions, we assess the perceived understandability of test cases, while the last question provides a more accurate indication of how much a participant understand the content of the test.}

\revised{As for the order of the test classes, half of the participants will first engage with the test class generated by \textsc{Mosa} and then with the one generated by \boost. Conversely, the other half of the participants will read the two test classes in the reverse order.}

\revised{Through the experiment, we will be able to assess the extent to which developers can understand and deal with the information provided by the test cases generated by the two approaches.}

\subsection{\revised{Data Analysis}} 
\revised{Once computed all the metrics to address our four research questions, we plan to run statistical tests to verify whether the differences observed between \boost and \textsc{Mosa} are statistically significant. More specifically, we aim to employ the non-parametric Wilcoxon Rank Sum Test~\cite{conover1980practical} (with $\alpha == 0.05$) on the distributions of (1) code coverage, (2) mutation coverage, (3) size per test case, (4) weighted method count of a test suite, (5) efferent coupling, (6) number of test smells, and (7) understandability scores assigned by developers in the user study.
In this respect, we formulate the following null hypotheses:}

\begin{description}[leftmargin=0.3cm]
	\item[Hn 1.] There is \emph{no significant difference} in terms of \emph{branch coverage} achieved by \boost and MOSA.
	
	\smallskip
	\item[Hn 2.] There is \emph{no significant difference} in terms of \emph{mutation coverage} achieved by \boost and \textsc{Mosa}.
	
	\smallskip
	\item[Hn 3.] There is \emph{no significant difference} in terms of \emph{size per unit} achieved by \boost and \textsc{Mosa}.
	
	\smallskip
	\item[Hn 4.] \revised{There is \emph{no significant difference} in terms of \emph{weighted method count of a test suite} achieved by \boost and \textsc{Mosa}.}
    
    \smallskip
	\item[Hn 5.] \revised{There is \emph{no significant difference} in terms of \emph{efferent coupling} achieved by \boost and \textsc{Mosa}.}
	
	\smallskip
	\item[Hn 6.] \revised{There is \emph{no significant difference} in terms of the \emph{number of test smells} achieved by \boost and \textsc{Mosa}.}	
	
	\smallskip
	\item[Hn 7.] \revised{There is \emph{no significant difference} in terms of the \emph{understandability scores} achieved by \boost and \textsc{Mosa}.}	
	
\end{description}

From a statistical perspective, we have to take into account the fact that, if one of the null hypothesis is rejected, then one between \boost and \textsc{Mosa} is statistically better than the other. Hence, we defined a set of alternative hypotheses such as the following: 

\begin{description}[leftmargin=0.3cm]
	\item[An 1.] The \emph{branch coverage} achieved by \boost and MOSA is \emph{statistically different}.
		
	\smallskip
	\item[An 2.] The \emph{mutation coverage} achieved by \boost and \textsc{Mosa} is \emph{statistically different}.
		
	\smallskip
	\item[An 3.] The \emph{size per unit} of the unit test suites generated by \boost and \textsc{Mosa} is \emph{statistically different}.
	
	\smallskip
	\item[An 4.] \revised{The \emph{weighted method count of a test suite} of the unit test suites generated by \boost and \textsc{Mosa} is \emph{statistically different}.}
    
    \smallskip
	\item[An 5.] \revised{The \emph{efferent coupling} of the unit test suites generated by \boost and \textsc{Mosa} is \emph{statistically different}.}
	
	\smallskip
	\item[An 6.] \revised{The \emph{number of test smells} of the unit test suites generated by \boost and \textsc{Mosa} is \emph{statistically different}.}
	
	\smallskip
	\item[An 7.] \revised{The \emph{understandability scores} of the unit test suites generated by \boost and \textsc{Mosa} is \emph{statistically different}.}
	
\end{description}

We reject the null hypotheses if $Hn_i \iff p < 0.05$. In addition to the Wilcoxon Rank Sum Test, we rely on the Vargha-Delaney ($\hat{A}_{12}$)~\cite{van2001refactoring} statistical test to measure the magnitude of the differences in the distributions of the considered metrics. Based on the direction given by $\hat{A}_{12}$, we can make a practical sense to the alternative hypotheses. Should the $\hat{A}_{12}$ values be lower than 0.5, this would denote that the test suites generated by \boost would be better than those provided by \textsc{Mosa}. For instance, a  $\hat{A}_{12} < 0.50$ in the distribution of code coverage would indicate that the code coverage achieved by \boost is higher than the one reached by the baseline. Similarly, a $\hat{A}_{12} > 0.50$ indicates the opposite, while $\hat{A}_{12} == 0.50$ points out that the results are identical.

\revised{Besides the statistical analysis of the distributions collected in our empirical study, we will also proceed with the verification of the assertion messages written by the user study participants. In this case, the first two authors of the paper will act as \emph{inspectors} and assess whether the reported messages are in line with the actual behavior of the test cases. In doing so, the inspectors will take advantage of a code coverage analysis tool, which might reveal the path covered by a test and, therefore, help assessing the match between the messages and the goals of the test case. We will finally collect and report the number of matches between assertion messages and actual behavior of the tests for each of the experimented tools. In addition, we will make use of the free answers provided by participants when explaining the reasons for the understandability score (question \#2 of the task) to identify the reasons for the correct/wrong assertion definitions. We will hence be able to provide an overview of the advantages and disadvantages of each test case generation tool with respect to the understandability of the resulting test cases.}

\subsection{Publication of generated data}
\boost source code, as well as all the other data generated from our study will be publicly available in an online repository (e.g., \textsc{GitHub}). \revised{We also plan to release the scripts to automatically generate the test suites, other than the data collected and used for the statistical and content analysis that we will present in the paper.}
    \section{Limitations}
\label{sec:limitations}
This section discusses the main limitations of the study.

A first possible limitation could be connected to the selection of the baseline technique on which we build \boost. The selection of \textsc{Mosa} is driven by the fact that this is the technique we know best and feel most confident with to modify. Yet, we believe that the selection of another baseline do not have an important impact on the results obtained in the context of our study. In particular, our aim is to define a systematic approach and to improve the resulting structure of the generated test cases, independently from the baseline approach, \ie the methodology implemented in \boost can be applied on any automatic test case generation technique. In any case, we also plan to replicate the study with different core techniques in the future, in order to verify this consideration.


Other possible limitations could be related to the experimental design discussed in Section \ref{sec:emp-study}. First, it is worth to note that the parameters used for the algorithms' configuration can influence the outcomes. To this aim we will rely on the default settings available in \textsc{Evosuite} on the basis of previous research in the field \cite{arcuri2013parameters} which showed that the configuration of parameters is not only expensive but also possibly ineffective in improving the performance of search-based algorithms. Moreover, since our approach and the baseline selected for comparison are both implemented within the same tool, \ie \textsc{Evosuite} \cite{evosuite}, we will rely on exactly the same the same underlying implementation of the genetic operators, avoiding possible confounding effects due to the use of different algorithms. 

To deal with the inherent randomness of genetic algorithms, we are going to re-execute our experimental procedure 30 times---as recommended by previous research \cite{campos2017empirical}---and report their average performance when discussing the results. 

Finally, we decided to adopt well-known state-of-the-art metrics for comparing techniques' performance. For example, we select branch coverage to assess the effectiveness of the tests generated by the two approaches. 
In addition, we employ appropriate statistical tests to verify the significance of the differences achieved by our approach and the baseline. Specifically, we rely on the Wilcoxon Rank Sum Test~\cite{conover1980practical} for statistical significance and the Vargha-Delaney effect size statistic~\cite{van2001refactoring} to estimate the magnitude of the observed difference. 

\revised{Last but not least, in the context of the user study conducted to assess the understandability of the generated test cases, we recruit participants using a research-oriented platform like \textsc{Prolific}. While this choice might potentially introduce some sort of selection bias \cite{reid2022_prolific_recommendations}, the platform allows to recruit participants from all around the world, with different expertise and experience. Whenever needed, we will filter out responses of participants who did not take the task seriously, other than recruit additional participants to reach an acceptable amount. Other than collecting background information by directly inquiring participants, the online platform that we will use can keep track of the time spent by each participant on each answer: this will enable an improved analysis of the performance of the participants, possibly helping us spot cases to discard. Nonetheless, we are aware of the limitations of an online experiment - yet, in the current pandemic situation, this is the only viable solution.}
    \section{Conclusion}
\label{sec:conclusions}

The ultimate goal of our research is to define a systematic strategy for the automatic generation of test code. In this paper, we will start working toward this goal by implementing the concepts of intra-method and intra-class testing within a state-of-the-art automatic technique for test case generation like \textsc{Mosa}. One of the risks connected to these mechanisms is the decrease of effectiveness: by forcing our approach to generate intra-method tests we naturally limit its scope, potentially lowering the number of tangentially covered branches. 

The technique we propose is also prepared to allow the generation at different granularity levels. Indeed, one can simply increase the number of production calls allowed in the first part of the generation, that we limit to one in this first concept, to generate tests at incremental levels of granularity. This would potentially have key implications, as the proposed strategy can be easily extended from a two-step (\ie \emph{intra-method} + \emph{intra-class}) to an n-step approach in which the number of calls allowed to methods of the class under test (CUT) is increased at each step. Since different number of calls to methods of the class under test corresponds to different paths on the state machine of the CUT, it would be possible to limit the length of the paths to execute on the state machine, thus providing shorter and more comprehensible tests for which it will be easier to generate an oracle. In this sense, our work poses the basis for the definition of a brand new way to generate test cases that might be of particular interest for the researchers working at intersection between software testing and software code quality. 



As part of our future work, we plan to assess our technique following the methodology described in Section \ref{sec:emp-study}. Later, we plan to exploit the granular nature of \boost to perform an experimentation based on several granularity levels. Finally, we plan to implement our approach on top of a broader set of baselines techniques as well as an \emph{in-vivo} performance assessment involving real testing experts.
	
	
%
%
%

	\section*{Acknowledgements}
	Fabio is partially supported by the Swiss National Science Foundation through the SNF Project No. PZ00P2\_186090 (TED).

	\balance
	\bibliographystyle{ACM-Reference-Format}
	\bibliography{main}
\end{document}